# Outstanding $T_C$ Enhancement in 5*d*–3*d* $Y_2NiIrO_6$ by Compression

Zheng Deng[a,*], Yao Zhang[b], Sijia Zhang[a], Jing Song[a], Wanli He[c], Yuanzhe Li[c], Meilin Jin[c], Xiang Li[c], Guanghua Liu[d], Zhen Dong[d], Jinkai Bi[d], Wenmin Li[d], Jianfa Zhao[a], Jun Zhang[a], Yi Peng[a], Luchuan Shi[a], Junling Meng[b,*], Xiancheng Wang[a], Changqing Jin[a,*]

[a] *Beijing National Laboratory for Condensed Matter Physics, Institute of Physics, Chinese Academy of Sciences and School of Physics, University of Chinese Academy of Sciences, Beijing, 100190, China*

[b] *School of Chemistry, Jilin Normal University, Changchun, 130024, China*

[c] *Centre for Quantum Physics, Key Laboratory of Advanced Optoelectronic Quantum Architecture and Measurement (MOE) and School of Physics, Beijing Institute of Technology, Beijing, 100081, China*

[d] *Institute of Quantum Materials and Physics, Henan Academy of Sciences, Zhengzhou, 450046, China*

Correspondence to: Z. Deng (dengzheng@iphy.ac.cn), J. Meng (mengjunling@jlnu.edu.cn), C. Jin (jin@iphy.ac.cn)

**Abstract**

Understanding and predicting the properties of 5*d* compounds critically depend on the identification of the superexchange interactions from which their magnetism emerges. The study of pressure effects on double perovskite $Y_2NiIrO_6$ (YNIO) provides deep insight toward this goal. At ambient pressure, YNIO is a ferrimagnetic insulator with the $Ir^{4+}$-5*d* $J_{eff} = 1/2$ Mott-insulating state. Under physical pressure up to 17 GPa, the compound exhibits concurrent compression on Ni/Ir–O bond lengths and Ni–O–Ir bond angles, leading to an increase of the Curie temperature from 192 to 243 K. On the contrary, external pressure increases distanced Ir–Ir interaction and in turn induces magnetic frustration in $Sr_2IrO_4$/$Sr_3Ir_2O_7$ due to the extended 5*d* orbitals. In YNIO, the rock-salt ordered Ni–Ir naturally blocks extended superexchange beyond the nearest neighbor, and in turn suppresses such magnetic frustration. Moreover, the orthogonal Ni $e_g$–Ir $t_{2g}$ pathway in YNIO is robust under lattice distortion, while the superexchange is weakened by bond bending in $La_2NiMnO_6$ with a similar half-filed $e_g$–$t_{2g}$ configuration. Our findings establish a framework for elucidating the mechanism of 5*d*–3*d* superexchange and guide bond-engineered magnetism in iridate-related systems.

## 1. Introduction

Iridates and other $5d$ oxides have received extensive interest due to the renewed perception distinguished from their $3d$ counterparts[1-4], due to the comparable spin-orbit coupling (SOC), Coulomb repulsion, and crystal splitting field in the former. Although the more spatially extended $5d$ orbitals suggest a tendency toward metallic behavior, this traditional picture is contradicted in systems such as the Ruddlesden-Popper iridates $Sr_2IrO_4$ and $Sr_3Ir_2O_7$[5]. These perovskite-related phases are antiferromagnetic insulators with high ordering temperatures but small magnetic moments. This exotic state arises from the $t_{2g}$ orbitals splitting of $Ir^{4+}$ $5d^5$ into a lower fourfold ($J_{\mathrm{eff}} = 3/2$) and an upper twofold ($J_{\mathrm{eff}} = 1/2$) level by the strong SOC. The half-filled $J_{\mathrm{eff}} = 1/2$ doublet is further split by the onsite Coulomb repulsion, giving rise to a Mott insulating state. Despite extensive studies, a couple of puzzles remain unresolved, including possible hidden magnetic states or properties due to the extremely small magnetic moments of Ir, and the persistent insulating states under high pressures[5-7].

The $5d$–$3d$ hybrid lattice offers a promising route to amplify and probe magnetic states of $5d$ via exchange coupling with strong magnetic and localized $3d$ electrons[8-11]. The double perovskite ($A_2B'B''O_6$) structure provides a natural platform for such combinations because the two distinct crystallographic B-sites can be flexibly occupied by cations ranging from $3d$ to $5d$ series[12, 13], and the transport and magnetic properties can be tuned by lattice distortion via independent A-site substitution[14-18]. A prototypical example is $Y_2NiIrO_6$ (YNIO), where $Ni^{2+}$ and $Ir^{4+}$ are strongly coupled in a long-range ferrimagnetic (FiM) order with Curie temperature ($T_C$) of 192 K[19]. Furthermore, this compound also hosts a $J_{\mathrm{eff}} = 1/2$ Mott-insulating state like the "pure" iridates. Fermi level of YNIO is dominated by Ir $5d^5$ orbitals, and the strong SOC along with nearly canonical octahedral crystal field lead to the splitting of the $J_{eff} = 1/2$ state, as confirmed by our and previous DFT studies[20]. Notably, chemical pressure effects on YNIO reveal that the band gap ($E_g$) is largely unaffected by octahedral tilting angle. On the other hand, the octahedral tilting is believed to account for the avoidance of pressure driven metallization in $Sr_2IrO_4$, $Sr_3Ir_2O_7$, etc[5-7, 21].

To completely understand the magnetism of the 5*d* electrons, one must clarify the 5*d*–3*d* magnetic interactions beyond the Goodenough-Kanamori-Anderson rules. Pressure is a powerful method for probing superexchange mechanism by exploring the relationship between crystal structure, electronic bands, and the magnetic properties. In this work, we focus on the evolution of YNIO magnetic properties under high-pressure. YNIO shows distinct pressure effects from 5*d* $Sr_2IrO_4$/$Sr_3Ir_2O_7$ and its isoelectronic counterpart $La_2NiMnO_6$, owing to its unique orbital configuration. Furthermore, through the compelling comparison with chemical compressed $Lu_2NiIrO_6$[22, 23], we demonstrate that the large enhancement of 5*d*–3*d* superexchange and reduction of band gap originates from increased Ni– and Ir–O orbital hybridization, which is predominantly facilitated by the shortening of the Ni–O and Ir–O bond lengths rather than the bending of Ni–O–Ir angles. These results also exhibit the exotic exchange-bias effect and large coercivity of YNIO could be exploited for spin valves and reliable data storage as its magnetic ordering temperature is raised toward room temperature[19].

## 2. Results

### 2.1 Pressure-Induced Structural Evolution

At ambient pressure, YNIO crystallizes in a *B*-site rock-salt ordered monoclinic structure with space group of $P2_1/n$ (Figure 1a). Ni and Ir occupy 2*c* (0.5, 0, 0.5) and 2*d* (0.5, 0, 0) respectively, with Ni-Ir antisite ratio of 7.5%. This low antisite ratio is an advantage for clarifying the Ni–Ir superexchange. The structural evolution of YNIO under high pressures was investigated at room temperature using *in-situ* synchrotron X-ray diffraction (XRD) up to 35 GPa. Figure S1 displays a systematic shift of all Bragg peaks to higher angles with increasing pressure, and all patterns remained indexable with the monoclinic symmetry remaining stable up to 26 GPa. Thus, we performed the Rietveld refinements on these XRD pattern using the ambient structure model (Figure 1b, 1c, Figure S2, and Table S1)[24]. The pressure dependence of the unit cell volume (*V*-*P*) up to 25 GPa in Figure 1d exhibits monotonic compression, with a noticeable anomaly near 17 GPa suggestive of a pressure-driven structural phase transition. This transition is further evidenced by the *in-situ* Raman spectrum (Figure S3), where one can find a new band at ~600 $cm^{-1}$, indicating the structural phase transition at around 17 GPa. To quantify the compressibility, the *V*-*P* curve of the low-pressure phase is fitted with the third-order Birch-Murnaghan equation of state[25]:

$$P = \frac{3}{2}B_0\left[(\frac{V_0}{V})^{7/3} - (\frac{V_0}{V})^{5/3}\right] \times \left\{1 + \frac{3}{4}(B_0' - 4) \times \left[(\frac{V_0}{V})^{2/3} - 1\right]\right\}$$

The fitting for the low-pressure phase yields the bulk modulus $B_0$ =300(4) GPa and its derivative $B_0'$ = 3.8(1). For comparison, the chemically compressed analogue $Lu_2NiIrO_6$ has a lattice volume of 221.5 Å$^3$, close to YNIO at around 8 GPa (Figure 1d)[22]. Note that the analysis of the high-pressure phase is beyond the scope of this work and will be discussed in a separate publication.

Figure 1e summarizes the pressure dependence of the lattice constants, revealing anisotropic compression. Given the nearly linear decrease of the lattice constants with pressure, the axial compressibility $\kappa$ is defined as $\kappa_a = -\frac{1}{a}\frac{\Delta a}{\Delta P}$. The obtained $\kappa$ for YNIO and the reported value of other monoclinic double perovskites are presented in Table 1. YNIO exhibits remarkable anisotropic axial compressibility, with the *b*-axis being exceptionally incompressible. Although serval theoretical studies have been conducted on YNIO and its analogues, such as $Ba_2NiIrO_6$ and $Lu_2NiIrO_6$[26, 27], corresponding experimental results have yet to be obtained. Further investigation incorporating more subtle parameters, such as local octahedral distortions, strong SOC, and possible Ir orbital ordering, could be needed to clarify this anomalous behavior.

**Table 1.** Experimental axial compressibility of reported monoclinic double perovskite materials to date. The unit of $\kappa$ is $10^{-3}$ $GPa^{-1}$. * Estimated from the figures in the corresponding references. # Measured at 10 K.

| compound | space group | $\kappa_a$ | $\kappa_b$ | $\kappa_c$ | $\kappa_b/\kappa_a$ | $\kappa_b/\kappa_c$ | $\kappa_a/\kappa_c$ |
|---|---|---|---|---|---|---|---|
| YNIO | $P2_1/n$ | 1.53 | 0.27 | 1.06 | 0.18 | 0.25 | 1.44 |
| $Sr_2ZnWO_6$*[25] | $P2_1/n$ | 2.71 | 2.02 | 1.44 | 0.75 | 1.4 | 1.88 |
| $Sr_2CoWO_6$[28] | $P2_1/n$ | 1.60 | 1.41 | 0.99 | 0.88 | 1.42 | 1.62 |
| $Sr_2ZnIrO_6$[29] | $P2_1/n$ | 2.50 | 0.80 | 3.00 | 0.32 | 0.27 | 0.83 |
| $Ba_2BiTaO_6$*[30] | $I2/m$ | 2.68 | 2.86 | 1.32 | 1.07 | 2.17 | 2.03 |
| $Ca_{1.5}La_{0.5}FeMoO_6$#[31] | $P2_1/n$ | 1.76 | 4.1 | 4.47 | 2.33 | 0.92 | 0.40 |
| $Ba_2PrRu_{0.8}Ir_{0.2}O_6$*[32] | $P2_1/n$ | 0.82 | 0.81 | 1.45 | 0.99 | 0.56 | 0.57 |
| $La_2LiRuO_6$[33] | $P2_1/n$ | 1.70 | 2.41 | 3.98 | 1.42 | 0.61 | 0.43 |

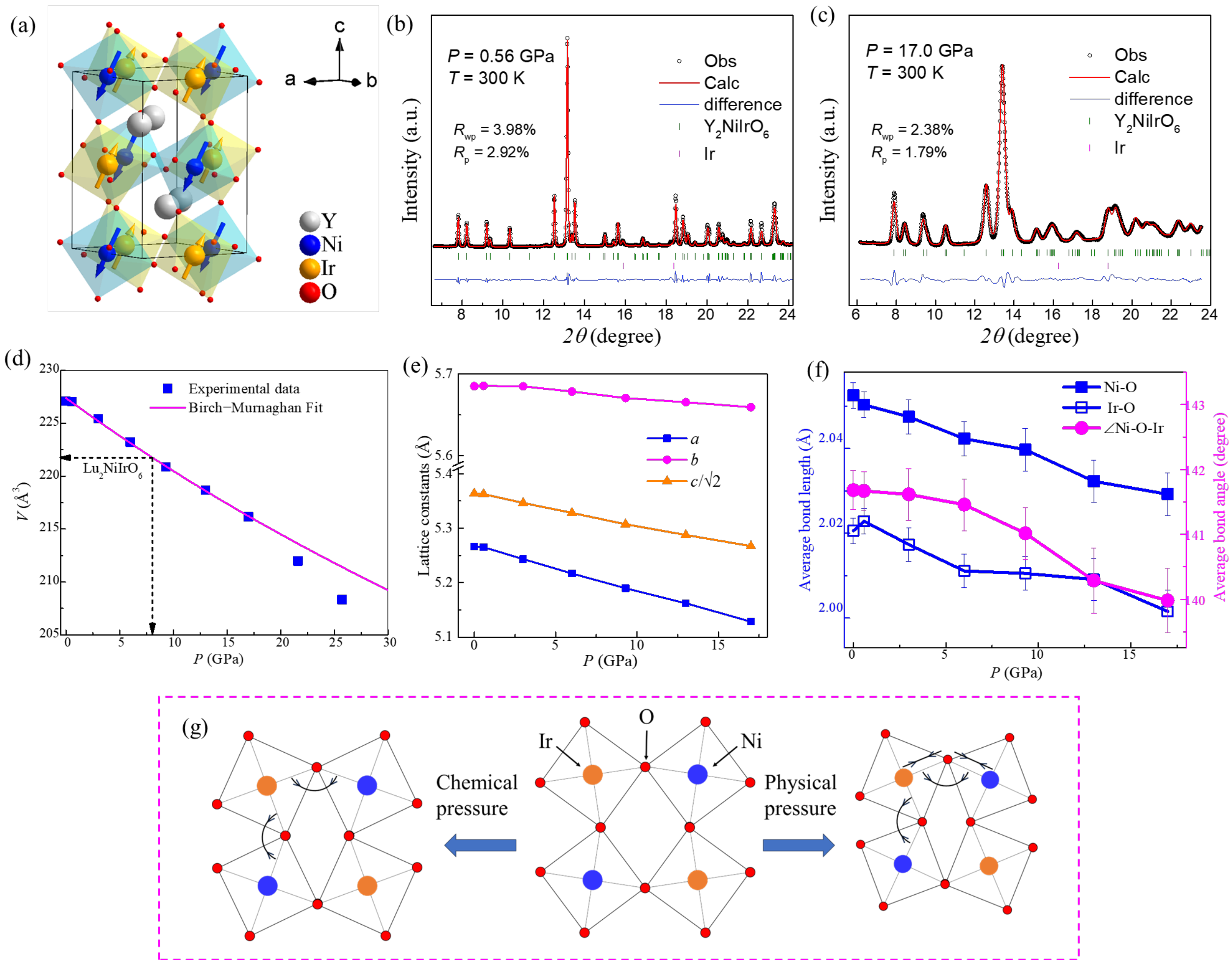

**Figure 1. The changes in the crystal structure of YNIO under pressures.** (a) Schematic crystal structure and magnetic structure of YNIO at ambient pressure. Synchrotron XRD pattern along with the Rietveld refinement at (b) 0.56 and (c) 17.0 GPa. (d) Cell volume as a function of pressure and the fit to Birch-Murnaghan equation of state below 17 GPa. The arrows indicate the cell volume of $Lu_2NiIrO_6$ and its corresponding chemical pressure. Pressure dependences of (e) lattice constants, and (f) average Ni–O, Ir–O length, and ∠Ni–O–Ir angle. (g) Schematic octahedral distortions and lattice shrinkage after chemical and physical compression on YNIO.

It is worth noting that the low sensitivity of oxygen to XRD results in relatively large error bars for the position of the three O atoms. Nevertheless, the change trends for average Ni–O and Ir–O bond lengths, and average ∠Ni–O–Ir are still clear in Figure 1f. The total reductions of Ni–O, Ir–O, and ∠Ni–O–Ir are approximately 1% up to 17 GPa, roughly half of the compression rates of the *a*- and *c*-axes. In the chemically compressed analogue $Lu_2NiIrO_6$, the average ∠Ni–O–Ir = 139.6° is apparently larger than that of YNIO at 8 GPa (141.2°)[22]. This difference can be attributed to the origin of lattice compression by chemical substitution in this system. The identical trivalence of A-site cations ($Y^{3+}$ and $Lu^{3+}$) preserves $Ni^{2+}$ and $Ir^{4+}$ oxidation states, and both *B*-site cations retain the $BO_6$ octahedral coordination. As a result, Ni–O and Ir–O lengths should remain nearly unchanged[34]. Instead, the lattice compression by chemical pressures in the $Ln_2NiIrO_6$ family primarily leads to tilting of the $NiO_6$ and $IrO_6$ octahedra, *i.e.* the bending of ∠Ni–O–Ir. On the other hand, the physical compression on YNIO results in two concurrent effects: the octahedron tilting (bending of ∠Ni–O–Ir) and the direct octahedron shrinkage (shortening of the Ni–O and Ir–O bond lengths). The compelling comparison between the physical and chemical pressure effects on structural properties of $Ln_2NiIrO_6$ (*Ln* = Sm to Lu, and Y) is illustrated in Figure 1g.

**2.2 Band Gap Narrowing and Enhancement of Curie Temperature under Pressure**

The impact of shortened *B*-O bond length on band structure is clearly demonstrated in electronic transport measurements. Figure 2a shows the temperature dependence of resistance $R(1/T)$ under various pressures. While the overall resistance decreases with increasing pressure, YNIO retains semiconducting behavior across the entire measuring temperature range. All the $R(1/T)$ curves can be well fitted by the thermal activation model. The pressure dependence of $E_g$, which is derived from two independent experimental runs, is summarized in Figure 2b. The value of $E_g$ decreases monotonically from 0.36 eV at 0.6 GPa to 0.26 eV at 16.8 GPa, corresponding to an average rate of approximately –0.006 eV/GPa. On the other hand, the chemically compressed $Lu_2NiIrO_6$ exhibits semiconducting behavior with $E_g$ of 0.39 eV[22], exhibiting no noticeable reduction of $E_g$.

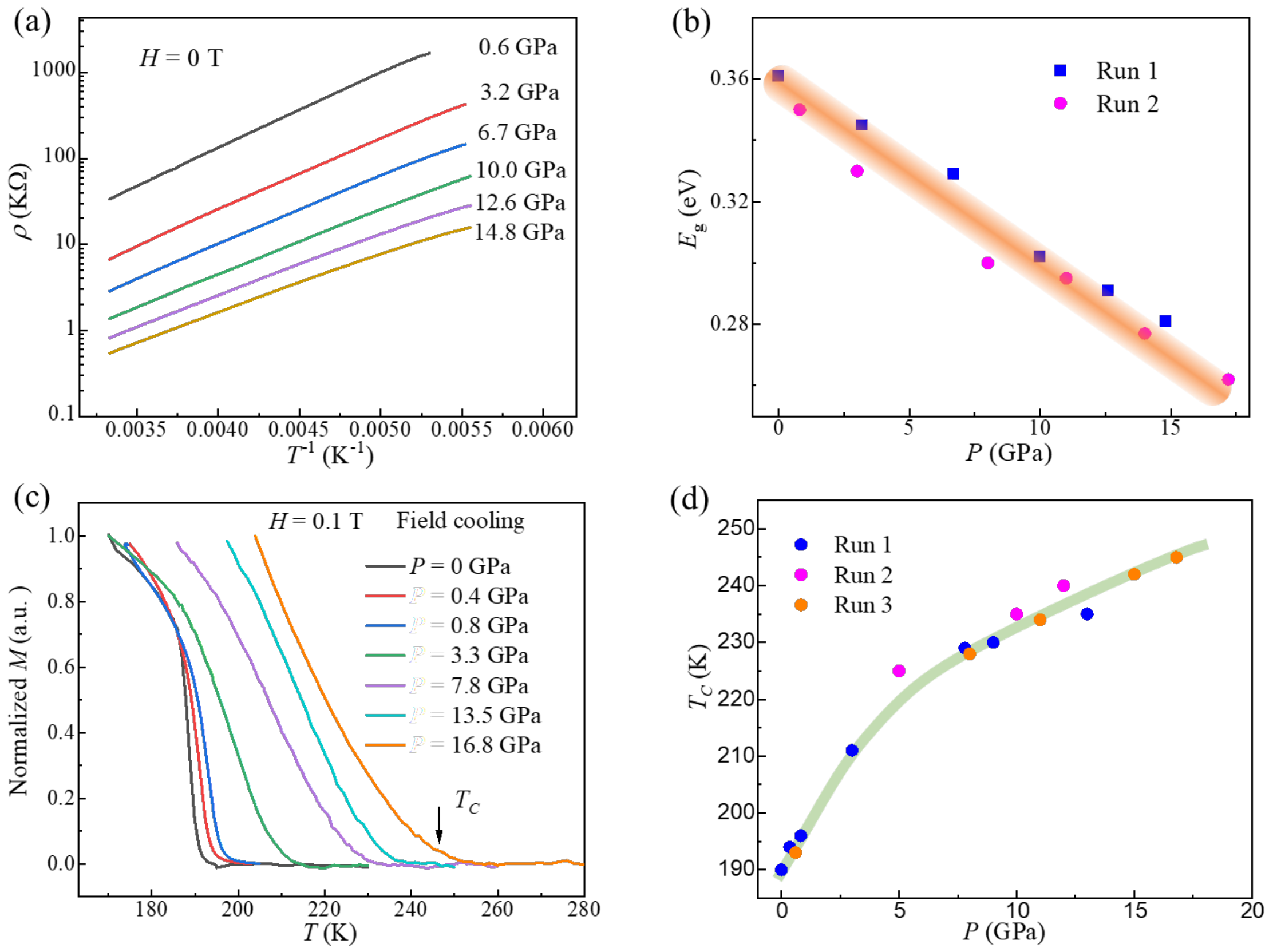

**Figure 2. The changes in electronic conduction and magnetic properties of YNIO under pressure.** (a) Resistance versus $T^{-1}$ and (b) $E_g$ under varying pressures of YNIO under pressures. (c) normalized $M(T)$ of YNIO at varying pressures. (d) $T_C$ of YNIO as a function of pressure.

At ambient pressure, YNIO exhibits a long-range FiM order with $T_C$ of 192K. Figure 2c shows the temperature-dependent magnetization ($M(T)$) under increasing pressure up to 16.8 GPa after the field cooling process in an external field of 0.1 T. Due to the limited sample quantity in the DAC, the magnetization data were background-subtracted using signals from an empty cell, and therefore the $M(T)$ curves in Figure 2c are normalized. The corresponding raw data are plotted in Figure S4. The broadenings of magnetic transitions at high pressures are attributable to the expected pressure gradient within the DAC.

$T_C$ is determined from the upturn on each $M$(T) curve. A significant increase in $T_C$ with compression is observed. Based on magnetic data from three independent experimental runs, the pressure dependence of $T_C$ is summarized in Figure 2d. The $T_C$-$P$ relationship exhibits a distinct change in slope around 5 GPa, with a rate of approximately +6 K/GPa at lower pressures and about +2 K/GPa at higher pressures. The maximum $T_C$ reaches ~243 K at around 17 GPa.

It is worth noting that although $T_C$ is enhanced by pressure, the net magnetization decreases, as shown in Figure S4. Under compression over 17 GPa may render $T_C$ experimentally invisible, even if it persists. Determining whether ferrimagnetic order survives in the high-pressure phase will necessitate future investigations using *in-situ* X-ray magnetic circular dichroism or neutron diffraction. This reduction in local moments upon compression can be ascribed to the enhanced hybridization between O and transition metals, which in turn promotes delocalization of $d$ electrons[35]. For the slope change in $T_C$ *vs*. pressure curve at 5 GPa, there is not anomaly behavior consistent with lattice constants or electrical resistance. This slope change is likely attributed to more subtle effects. A plausible scenario is the competition between pressure-enhanced superexchange and the simultaneous reduction of local magnetic moments.

**2.3 First-principles Analysis of Pressure Effects**

Our first-principles calculations successfully reproduced the evolution of crystal structure under physical pressures which are applied by compressing the cell volume of YNIO. As shown in Figure S6a, the lattice constants decrease anisotropically with increasing pressure. The compression rates of $a$- and $c$-axes are significantly larger than that of $b$-axis. Figure S6b shows that Ni/Ir–O lengths and ∠Ni–O–Ir decrease upon compression, in accordance with the experimental results. It is worth noting that most of the previous calculations on $Ln_2NiIrO_6$ family overlooked the role of Ni–O and Ir–O lengths[20, 36]. The partial density of states (PDOS) indicates that the valence band top and conduction band bottom are predominantly composed of the Ir-5*d* and O-2*p* orbitals. Specifically, $t_{2g}$ of Ir-5*d* splits into $J_{eff}$ = 3/2 and 1/2 states. Coulomb repulsion further splits the $J_{eff}$ = 1/2 into lower and upper Hubbard bands, resulting in a Mott-insulating ground state with $E_g$ of about 0.35 eV (spin up)[20, 36]. After compression, bandwidth increases and $E_g$ decreases to ~0.15 eV (spin up) at 90% of equilibrium volume ($V_{eq}$) as shown in Figure S6c. This reduction of $E_g$ is qualitatively consistent with the experimental results.

To clarify the magnetic properties under physical pressure, local moments on two B-sites magnetic cations and magnetic interactions between them were calculated. Figure 3a displays slight reduction in both Ni and Ir magnetic moments with increasing pressures. This decrease is resulted from the delocalization of *d* electrons, owing to the enhanced hybridization between O and Ni/Ir orbitals. Nevertheless, the net magnetization remains approximately constant due to the antiparallel alignment of the Ni and Ir moments. Given that the magnetic interactions in YNIO are dominated by nearest Ni–O–Ir AFM superexchange[20, 36], we evaluated the evolution of the exchange parameter ($J_{Ni\text{-}Ir}$) to indicate the change of $T_C$ under pressures. As shown in Figure 3b, monotonic increase of $J_{Ni\text{-}Ir}$ below 18 GPa successfully reproduces the experimental enhancement of $T_C$. In short, our DFT calculations evidence that enhanced hybridization between O and Ni/Ir orbitals accounts for the strengthened superexchange and the improved $T_C$ under physical pressures.

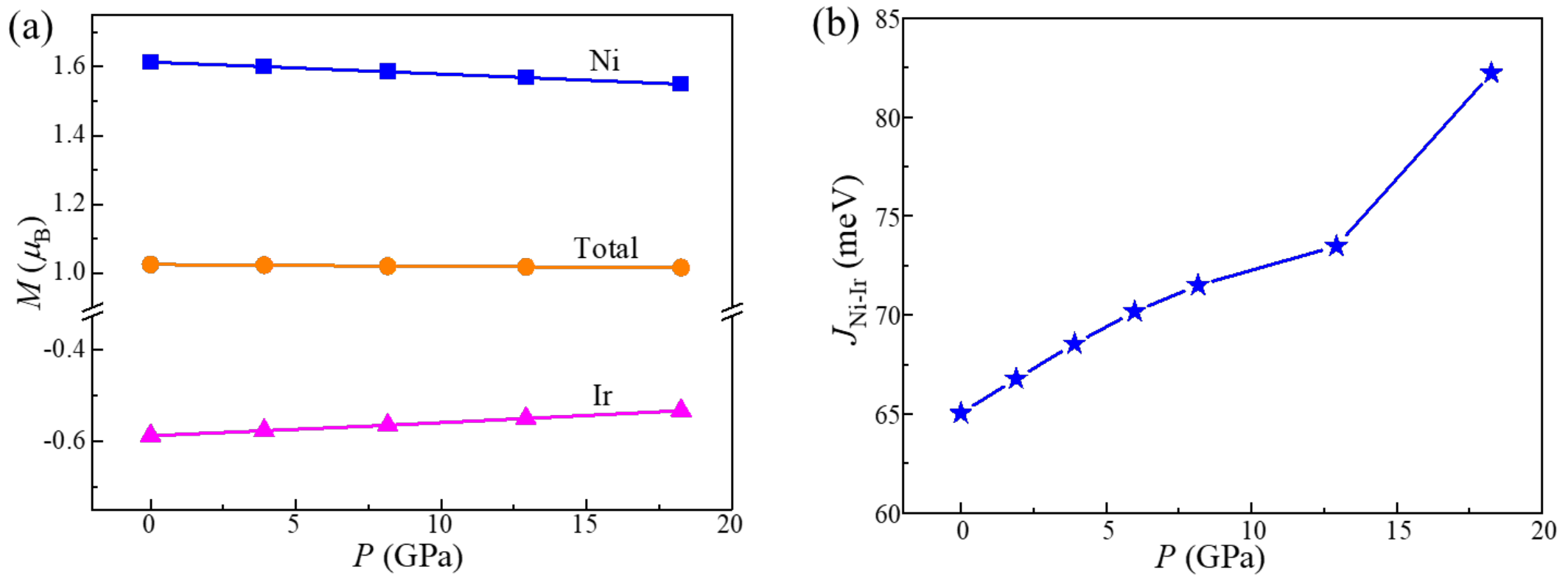

**Figure 3. Calculated magnetic properties of YNIO.** (a) The magnetic moments on Ni, Ir, and the net moments. (b) exchange parameter of nearest Ni and Ir, $J_{Ni\text{-}Ir}$, as functions of pressure.

## 3. Discussion

To elucidate the magnetic interactions in YNIO, we propose a model incorporating strong SOC on $Ir^{4+}$. According to Hund's rule, $3d^8$ of $Ni^{2+}$ yields a filled $t_{2g}$ and half-filled $e_g$ orbital. Since the filled $t_{2g}$ contributes negligibly to superexchange, the primary hopping pathways are simplified, focusing on the half-filled $e_g$ orbitals[37]. For the $5d^5$ of $Ir^{4+}$, the strong SOC splits the $t_{2g}$ into a $J_{eff}$ = 3/2 and $J_{eff}$ = 1/2 state, with the latter forming a half-filled Hubbard state near the Fermi level, as confirmed by our first-principles calculations.

The Ni−Ir superexchange can be established through two competing hopping paths as illustrated in Figure 4a, (i) AFM between the half-filled Ni $e_g$ and the half-filled Ir $J_{eff}$ = 1/2 in $t_{2g}$, (ii) FM between the half-filled Ni $e_g$ and the empty Ir $e_g$[37]. Although the Ni $e_g$ and Ir $t_{2g}$ orbitals are orthogonal in an ideal cubic double perovskite (∠Ni−O−Ir = 180°), the extended Ir-5$d$ orbitals and the distorted lattice with bended ∠Ni−O−Ir enable finite electron hopping[37, 38]. For the FM pathway, the $t_{2g}$–$e_g$ splitting in 5$d$ is several times larger than that of 3$d$, leading to the poor energetic overlap of their $e_g$ orbitals. However, we argue that the FM pathway is still not negligible, as evidenced by systems such as $Sr_2FeOsO_6$, which exhibits FM coupling along its *c*-axis via $3d\ e_g - 5d\ e_g$ overlap[39]. The $e_g$–$e_g$ overlap is highly sensitive to the bond angle, thus the FM interaction in YNIO is expected to be weak due to the buckled ∠Ni–O–Ir. As a result, the AFM superexchange dominates the Ni–Ir interactions in YNIO. Nevertheless, the residual $e_g$–$e_g$ hopping path is necessary for explaining the insensitivity of $E_g$ under chemical pressure, as discussed below.

For idea compression without structural distortions, pressure generally enhances superexchange in insulators[40, 41]. However, the distinct characters of 5*d* and lattice distortions lead to various pressure responses in numerous compounds. Because of the extended orbitals, magnetic interactions between first ($J_{1\text{Ir-Ir}}$), second ($J_{2\text{Ir-Ir}}$), and third ($J_{3\text{Ir-Ir}}$) neighboring Ir are relatively comparable in the prototypical 5*d* Mott insulator $Sr_2IrO_4$. Pressure enhances $J_{2\text{Ir-Ir}}$, $J_{3\text{Ir-Ir}}$ and in turn leads to a highly frustrated magnetic state. Consequently, the AFM order of $Sr_2IrO_4$ is gradually weakened by increasing external pressures[42], and similar frustrated magnetic state is also observed in $Sr_3Ir_2O_7$[43]. For double perovskite $La_2NiMnO_6$, which host a half-filed $e_g$-$t_{2g}$ configuration analogous to YNIO, its ferromagnetic ordering temperature shows minimal pressure dependence[44, 45]. The localized 3*d* electrons suppress orthogonal $e_g$–$t_{2g}$ hopping, whereas FM between the half-filled Ni $e_g$ and the empty $Mn^{4+}$ $e_g$ is dominated. Owing to the sensitivity of $e_g$–$e_g$ overlap to the bond angle, the buckling Ni-O-Mn bond angle counteracts the predicted increase of superexchange under pressure [44].

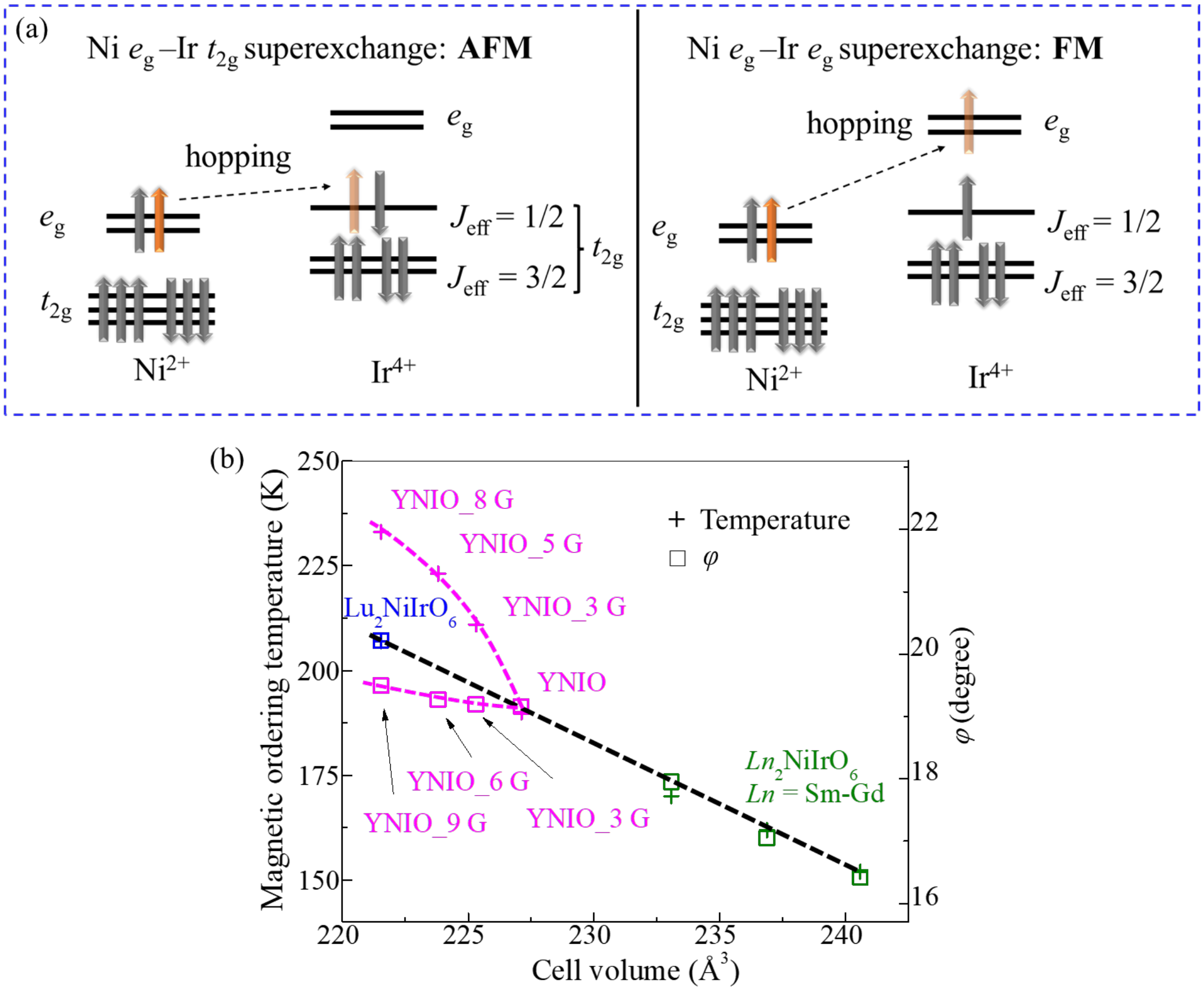

**Figure 4.** (a) Schematic of two superexchange pathways between Ni and Ir in YNIO. (b) Cell volume-dependent magnetic ordering temperatures and octahedral tilting angle of $Ln_2NiIrO_6$ at ambient pressure and YNIO at varying pressures.

On the other hand, YNIO occupies a unique intermediate regime. Its magnetic interaction is dominated by first neighboring Ni–Ir AFM superexchange[20, 36], avoiding the frustrated pressure-response in 5*d* system $Sr_2IrO_4$/$Sr_3Ir_2O_7$. The aforementioned robustness of the orthogonal AFM pathway against lattice distortion differs YNIO from the 3*d* double perovskite $La_2NiMnO_6$, in which the dominant FM $e_g$–$e_g$ path is diminished by bond angle bending. The major role of bond length compression in strengthening orbital hybridization and hopping explains why YNIO exhibits a pronounced enhancement of $T_C$ under pressure. This distinction underscores how structural tuning can be utilized to selectively enhance superexchange in multi-orbital materials.

A comparison between the physical and chemical compression on $Ln_2NiIrO_6$ ($Ln$ = Sm to Lu, and Y) (Figure 4b) indicates the dominant influence of bond shortening on Ni–Ir superchange. Under chemical compression (represented by the black dashed line), the magnetic ordering temperatures and octahedral tilting angles ($\varphi \equiv (180° - \langle\angle\text{Ni–O–Ir}\rangle)/2$) increase nearly linearly with decreasing lattice volume, *i.e.*, increasing chemical compression[19, 23, 38, 46-50]. Such consistency indicates that buckled ∠Ni–O–Ir strengthens the Ni–Ir AFM interaction. Chemical pressure on electronic transport and magnetic properties can be well demonstrated by the competing picture between AFM and FM paths. From YNIO to $Lu_2NiIrO_6$, the increased ∠Ni–O–Ir bending enhances Ni $e_g$–Ir $t_{2g}$ hybridization, thereby increasing $T_C$ from 192 to 207 K. However, such obvious orbital hybridization cannot account for the nearly unchanged $E_g$ with a single AFM path. Such inconsistent will be readily resolved when competition from FM path is considered. The bending ∠Ni–O–Ir dramatically decrease Ni $e_g$–Ir $e_g$ overlap because of its high sensitivity to the bond angle[39]. Consequently, the increased hopping in AFM coupling is effectively counterbalanced, in turn leaving $E_g$ nearly invariant. Turning to physical compression, the behaviors of YNIO clearly deviate from the linear trend defined by chemical compression (pink dashed lines, Figure 4b). The deviations can be exemplified by $Lu_2NiIrO_6$, with a lattice volume comparable to YNIO at round 8 GPa. $Lu_2NiIrO_6$ has a lower $T_C$ but larger octahedral tilting angle ($T_C$ = 207 K, and $\varphi$ = 20.2°/∠Ni–O–Ir = 139.6°) than the latter ($T_C \sim$ 230 K, and $\varphi$ = 19.5°/∠Ni–O–Ir = 141.0°). These divergences indicate a more effective mechanism beyond ∠Ni–O–Ir bending via physically compression. The fact that Ni–O, Ir–O, and ∠Ni–O–Ir in YNIO shrink at similar rates under physical compression (Figure 1f), highlights the dominant influence of bond shortening on orbital hybridization and electron hopping integrals. As a result, bandwidth increases and $E_g$ decreases steadily with increasing physical pressure, as confirmed by our DFT calculations.

**4. Conclusions**

In summary, we have performed comprehensive high-pressure studies on YNIO with the $J_{eff}$ = 1/2 Mott-insulating state. YNIO maintains insulating behavior from ambient pressure to 16.5 GPa, with $E_g$ decreasing from 0.36 to 0.26 eV. The physical pressure yields an enhancement in the $T_C$ from 192 to 243 K. The distinct pressure response on $Y_2NiIrO_6$ from iridates $Sr_2IrO_4$/$Sr_3Ir_2O_7$ and 3*d* double perovskites is mainly attributed to the alternate 5*d*-3*d* sublattices and the orthogonal 5*d* $t_{2g}$–3*d* $e_g$ pathway. This outstanding strengthened Ni-Ir superexchange is unambiguously a result of the direct shortening of Ni–O and Ir–O bonds, which is nearly inaccessible via chemical compression. Our findings highlight the distinct influences of bond length and bond angle on mediating magnetic interactions and suggest physical pressure as a powerful and clean method for tailoring magnetic properties in correlated compounds. We expect this work to stimulate further exploration of pressure- or film strain-engineered magnetism and deeper insight into the interplay between structure and superexchange in iridate-related systems.

**Methods**

Polycrystalline YNIO samples were synthesized under high-pressure and high-temperature conditions.[19] Laboratory X-ray diffraction (XRD) clarified the high purity and structural parameters at ambient pressure. The *in-situ* synchrotron XRD measurements were performed at beam-lines BL15U1 (wavelength = 0.6199

Å) of the Shanghai Synchrotron Radiation Facility (SSRF). Rietveld refinements were performed with GSAS software packages. The *in-situ dc* magnetization was measured with a Quantum Design Superconducting Quantum Interference Device (SQUID, MPMS3). The *in-situ* resistance measurements were conducted with a Quantum Design Physical Property Measurement System (PPMS). Below 1 GPa, samples were compressed with a Be-Cu alloy piston-cylinder cell for hydrostatic pressure. The pressures of the piston-cylinder cell were applied at room temperature and were measured by the superconducting temperatures of lead. Above 1 GPa, a Be-Cu diamond anvil cell (DAC) was used to produce higher pressures. It is worth noting that at low temperatures magnetization measurements with Be-Cu DAC is dominated by the signal from DAC itself (Figure S5 (c)). Because the Be-Cu alloy used in the DAC exhibits a large magnetic signal that is nearly proportional to the applied field. Consequently, obtaining sample signal by subtracting DAC background becomes unreliable and physically uninterpretable (Figure S5 (d)). The pressures of DAC were applied at room temperature and measured by the ruby fluorescence method.

All the first-principles calculations in this work were carried out by using the VASP (version 6.3.0) software package, which was developed based on the density functional theory[51]. The crystal structure of YNIO consists of 20 atoms. The interaction between valence electrons and ionic cores is dealt with by using the projector augmented-wave (PAW) method, while the exchange-correlation term is handled through the generalized gradient approximation (GGA) and the solid-state-corrected Perdew-Burke-Ernzerhof (PBEsol) method[52]. In the present calculations, we adopted the GGA+U method to calculate the ground-state electronic structure of YNIO, explicitly accounting for the electron correlation effects of the Ni-3*d* and Ir-5*d* shells. It is worth noting that higher-level methods, such as hybrid DFT functionals or DFT+DMFT, can provide improved quantitative accuracy for band gap predictions in Mott insulators. However, within the scope of the present study, the GGA+U framework provides an efficient and sufficient basis for our main conclusions. The Hubbard Coulomb potential was treated using the self-interaction correction, with $U_{eff} = U\text{-}J$, where $U$ and $J$ represent the Coulomb repulsion energy and the exchange-correlation energy, respectively. The $U_{eff}$ value of 4.0 eV was added to the Ni 3*d* orbitals with $U = 5.0$ eV and $J = 1.0$ eV. The $U_{eff}$ value of 2.6 eV was added to the Ir 5*d* orbitals with $U = 3.0$ eV and $J = 0.4$ eV. The effect of the $U_{eff}$ parameter was also examined, as presented in Table S3. Small variations in $U_{eff}$ exert negligible effects on the energy difference between the two magnetic configurations, and thus only marginally influence the magnetic coupling constant. The valence electron configuration O ($2s^2 2p^4$), Ni ($3d^8 4s^2$), and Ir ($5d^7 6s^2$) was calculated using the standard potential of the pseudopotential basis set. The entire Brillouin zone was inte$_{g}$rated using the 7×7×5 Monkhorst-Pack *k*-point sampling method. The plane wave basis set was used to expand the electron wave function, with a plane wave cutoff energy of 500 eV. The convergence criterion was set to a Hellman-Feynman force of ≤ 0.01 eV/A for each atom. The charge convergence criterion was set to < $10^{-4}$ eV. The tetrahedron method with Blöchl corrections for BZ inte$_{g}$ration was used after geometry optimization to obtain projected density of states (PDOS), refined energies, and the final charge density for Bader analysis.

**Acknowledgements**

The works at IOPCAS were supported by MOST of China (No. 2022YFA1403900, 2022YFA1403804, and 2024YFA1408003), NSF of China (No. 22531001), CAS Project for Young Scientists in Basic Research (No. YSBR-030), Beijing National Laboratory for Condensed Matter Physics (No.

2023BNLCMPKF006). The works at IQMP HNAS were supported by the High-level Talent Research Start-up Program Funding of Henan Academy of Sciences (No. 241827046, 242027151, and 241827022). Regarding the *in-situ* synchrotron XRD measurements, the beam-lines BL15U1 and BL17U1 of the Shanghai Synchrotron Radiation Facility (SSRF) are acknowledged.

**Ethics declaration**

The authors declare no competing financial interest.

**Data Availability Statement**

The data that support the findings of this study are available from the corresponding author upon reasonable request.

**Table of Contents**

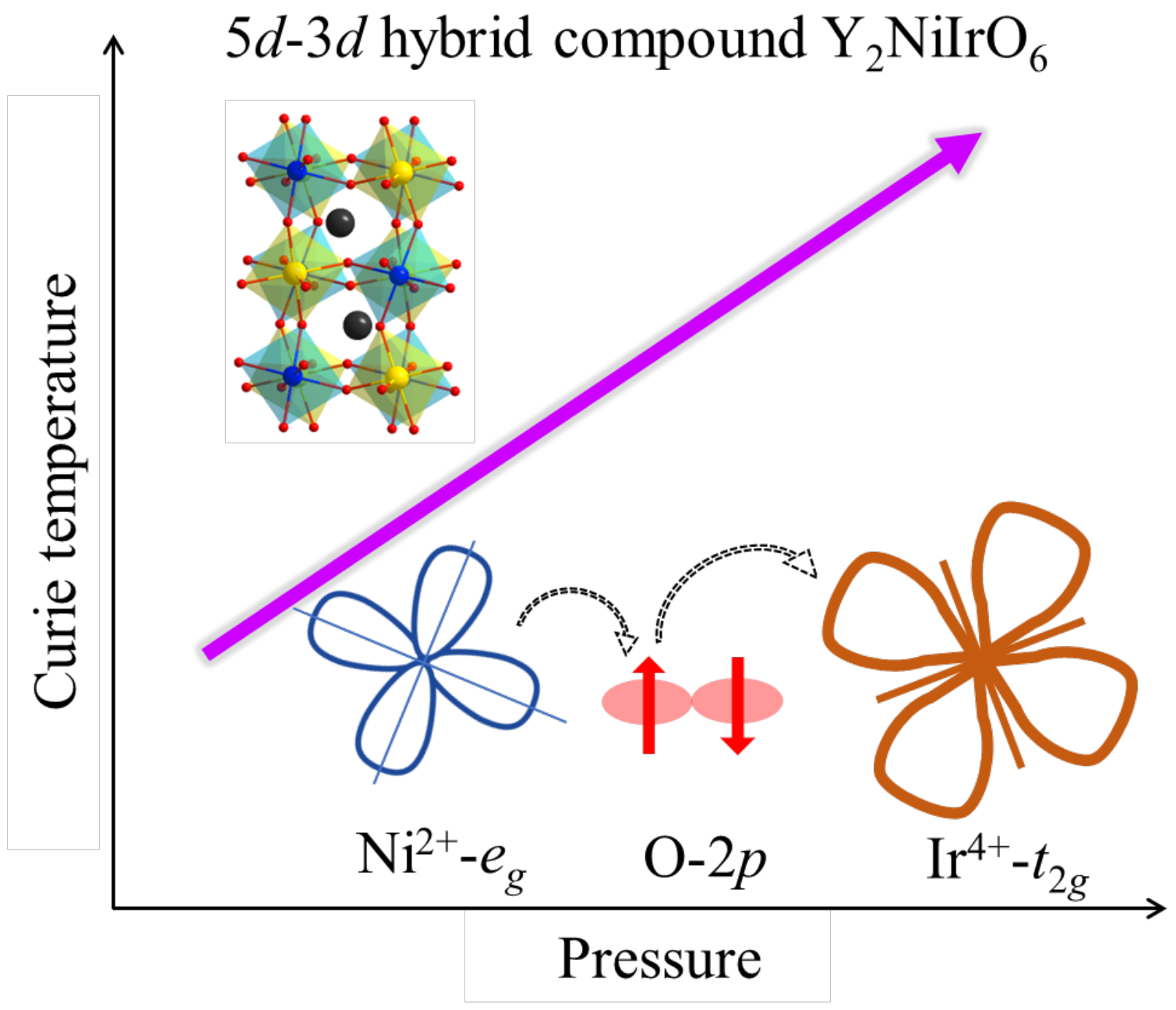

Double perovskite $Y_2NiIrO_6$ with half filed 3*d* $e_g$–5*d* $t_{2g}$ configuration, exhibits first experimental example of pressure-enhanced magnetic interaction in iridate and iridate-related systems. Its Curie temperature increase from 192 to 243 K at 17 GPa. Herein, the orthogonal $e_g$–$t_{2g}$ pathway remains robust even under compression-induced lattice distortion. This outstanding strengthened magnetic interaction is resulted from this robust superexchange pathway and direct shortened Ni/Ir–O bonds.